\documentclass[12pt]{article}

\usepackage[dvips]{epsfig}
\usepackage{amsfonts}

\makeatletter
\@addtoreset{equation}{section}%
\renewcommand{\theequation}{\thesection.\@arabic\c@equation}
\makeatother

\def\papertitlepage{\baselineskip 3.5ex \thispagestyle{empty}}
\def\preprinumber#1#2{\hfill \begin{minipage}{4.2cm}  #1
                 \par\noindent #2 \end{minipage}}

\begin{document}
\papertitlepage
\setcounter{page}{0}
\preprinumber{}{KEK-TH-1199}
\baselineskip 0.8cm
\vspace*{2.0cm}
\begin{center}
{\Large The Glueball Superpotential for $ G_2 $}
\end{center}
\vskip 4ex
\baselineskip 1.0cm
\begin{center}
 Osamu Saito \\
\vskip 1em
{\it High Energy Accelerator Research Organization (KEK)} \\
       \vskip -2ex {\it Tsukuba, Ibaraki 305-0801, Japan} \\
\end{center}
%%%%%%%%%%%%%%%%%%%%%%%
\vskip 5ex
%%%%%%%%%%%%%%%%%%%%%%%
%
\baselineskip=3.5ex
\begin{center} {\bf Abstract} \end{center}
\vskip 2ex
 We perform a perturbative computation of the glueball superpotential for ${\cal N}=1$ supersymmetric $G_2$ gauge theory
 with one adjoint matter multiplet. We find that the computation simplifies because the bosonic momentum integral cancels 
 with the fermionic momentum integral. The effective glueball superpotential allows us to gain an insight into non-perturbative aspect 
of the supersymmetric gauge theory.

\vspace*{\fill}
\noindent
November 2007

\newpage
\section{Introduction}
The non-perturbative aspects of supersymmetric gauge theories have been investigated for a long time 
\cite{Seiberg:1994rs,Seiberg:1994aj,Seiberg:1994pq}. Recently, Dijkgraaf and Vafa have pointed out that we can gain insight into 
non-perturbative gauge theoretic phenomena from a perturbative perspective\cite{Dijkgraaf:2002dh}: The relevant physical quantity
is the effective superpotential as a function  of the glueball superfield $S$, which is believed to behave as an elementary field
 in the IR.

%We perform a perturbative 
%computation of the effective superpotential as a function of the glueball superfield $S$, which is believed to behave as an 
%elementary field in the IR. Then we extremize the effective superpotential with respect to the glueball superfield $S$. 

In \cite{Dijkgraaf:2002xd}, the perturbative computation of the glueball superpotential was performed for $U(N)$ gauge theory
with one adjoint matter. The analysis was purely field-theoretic and diagrammatic. It was shown that the computation dramatically
 simplifies: only the {\it planar } diagrams contribute and furthermore the evaluation reduces to zero-dimensional field theory, 
i.e, a matrix model because of supersymmetry (for a review see \cite{Argurio:2003ym}).

Since the advent of \cite{Dijkgraaf:2002xd}, various attempts have been done to extend the diagrammatic analysis of 
\cite{Dijkgraaf:2002xd}. For example, multi-trace interactions and baryonic interactions were investigated in \cite
{Balasubramanian:2002tm} and in \cite{Argurio:2002hk}, respectively. The $U(N)$ gauged model with spontaneously broken 
${\cal N}=2$ supersymmetry was discussed in \cite{Itoyama:2007rr}. Other classical gauge groups were studied in \cite{Ita:2002kx}
and \cite{Kraus:2003jf}. For $SO(N)/Sp(N)$ with an adjoint matter, Feynman diagrams are associated to orientable and 
non-orientable Riemann surfaces. 

In this paper, we consider $G_2$ gauge theory with an adjoint matter. $G_2$ is different from classical gauge groups in that
it has an invariant three-tensor $f_{ijk}$. We derive Feynman rules and show that new types of diagrams appear. Then we compute
the effective glueball superpotential.

The paper is organized as follows: in section 2 we give a brief review of the perturbative  computation of the glueball
superpotential for classical gauge groups (in particular, $U(N)$ and $SO(N)$). In section 3 we investigate $G_2$ gauge group. 
Section 4 is devoted to conclusions and discussion. In appendix we summarize the properties of an invariant tensor $f_{ijk}$.

\section{Perturbative computation of glueball superpotentials}

In this section we show, following \cite{Dijkgraaf:2002xd}, how to compute the glueball superpotential. Let $\Phi$ be a massive
chiral superfield and belongs to some representation of the gauge group. The starting point is the following ${\cal N}=1$ four 
dimensional action
\begin{equation}
  S_{{\bf SYM}}(\Phi,\bar{\Phi})=\int d^4 x d^4 \theta \bar{\Phi}e^{V}\Phi +\int d^4 x d^2 \theta W(\Phi)+h.c,
\end{equation}  where $W(\Phi)$ is the gauge invariant superpotential. We use the conventions given in \cite{Gates:1983nr}.
The gauge field strength ${\cal W}_{\alpha}(=i\bar{D}^2e^{-V}D_{\alpha}e^{V})$ is treated as background. The glueball
superfield is defined by
\begin{equation}
  S=\frac{1}{32\pi^2}tr({\cal W}^{\alpha}{\cal W}_{\alpha}).
\end{equation}
We are looking for the
perturbative part of the effective superpotential of this system$\int d^2 \theta W^{pert.}_{eff}(S)$ as a function of the external
glueball superfield $S$. 

In \cite{Dijkgraaf:2002xd}, it is shown that for our purpose the relevant action can be written as
\begin{equation}
   S_{{\bf SYM}}(\Phi)=\int d^4 x d^2 \theta \left( -\frac{1}{\bar{m}}\Phi(\Box-i{\cal W}^{\alpha}D_{\alpha})\Phi+W_{\bf tree}
(\Phi) \right).
\label{action}
\end{equation}
In this derivation the anti-chiral superpotential is set to
\begin{equation}
  \bar{W}(\bar{\Phi})=\frac{1}{2}\bar{m}\bar{\Phi}^2
\end{equation}
and the anti-chiral superfield $\bar{\Phi}$ is integrated out.  Since the holomorphic quantity is independent of $\bar{m}$,
we can set $\bar{m}=1$. The expression (\ref{action}) is valid for any gauge groups. We discuss $U(N)$ theory in section 2.1 and 
$SO(N)$ in section 2.2.

\subsection{$U(N)$ gauge theory with an adjoint matter}

Let us consider $U(N)$ gauge theory interacting with a matter field $\Phi$ in the adjoint representation\cite{Dijkgraaf:2002xd}.
 In this case we see
 a dramatic simplification. To be specific we consider the following cubic superpotential
 \begin{equation}
   W_{tree}(\Phi)=\frac{m}{2}tr(\Phi^2)+\frac{g}{6}tr(\Phi^3).
   \label{cubic_superpotential}
 \end{equation}
 
\begin{figure}[htbp]
  \begin{center}
  \includegraphics{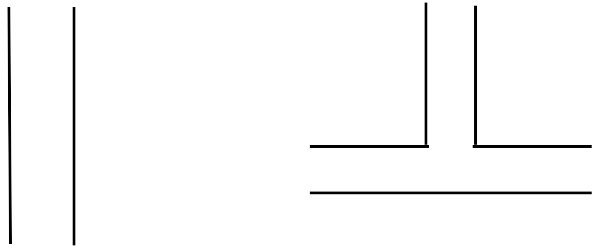}
  \caption{The propagator and the cubic vertex in terms of 't Hooft double line notation}
  \label{Feynman_U_N}
  \end{center}
\end{figure}

 The action (\ref{action}) and the mass term of the superpotential (\ref{cubic_superpotential}) lead to 
\begin{equation}
    \langle  \Phi \Phi \rangle = \frac{1}{p^2+m+{\cal W}^{\alpha}\pi_{\alpha}}
 \label{propagator_U_N}
\end{equation}   
for the momentum space propagator. $p$ denotes the four-dimensional bosonic and $\pi$ denotes the fermionic momentum. Since we
set $\bar{m}=1$, the expression (\ref{propagator_U_N}) is not manifestly dimensionally correct. 
Feynman diagrams corresponding to the propagator and cubic interaction can be represented as Fig.\ref{Feynman_U_N}.
We use the 't Hooft double line notation to keep track of the gauge index structure.

\begin{figure}[htbp]
  \begin{center}
    \includegraphics{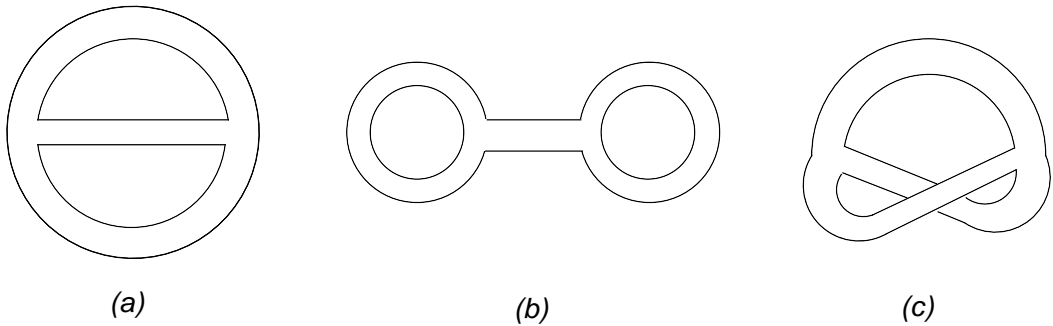}
    \caption{The vacuum diagrams of $U(N)$ gauge theory}
    \label{vacuum_diagram_U_N}
  \end{center}
\end{figure}

In order to compute the purely holomorphic part of the full partition function $Z^{\prime}$:
\begin{equation}
   Z^{\prime}=\int {\cal D}\Phi e^{-S_{\bf SYM}(\Phi)},
\end{equation} we have to evaluate vacuum diagrams  such as Fig.\ref{vacuum_diagram_U_N}.
In double line notation we associate each ribbon Feynman diagram to a two-dimensional surface(or Riemann surface). This is done
by filling in every index loop with a disk. For example diagram $(a)$ and $(b)$ are associated to $S^2$ while $(c)$ is 
associated to a torus $T^2$. In the case of $U(N)$ gauge theory with an adjoint matter, we obtain {\it orientable} and
 closed surfaces.
 
\begin{figure}[htbp]
   \begin{center}
     \includegraphics{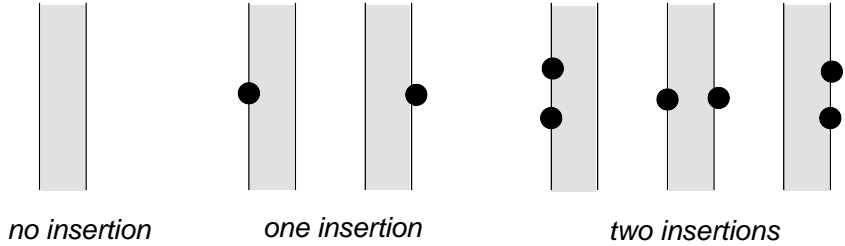}
     \caption{The propagator with insertions of the gauge field strength ${\cal W}^{\alpha}$. 
               The dots denote the insertions of ${\cal W}^{\alpha}$}
     \label{insertion_U_N}
   \end{center}
\end{figure}

 One of the remarkable fact is that only $S^2$ graphs (or planar graphs) contribute to the glueball superpotential. To show
 this we begin by investigating the propagator (\ref{propagator_U_N}) in detail. We introduce a Schwinger parameter $s$ and 
 expand the exponential with respect to the fermionic momentum:
\begin{eqnarray}
  \frac{1}{p^2+m+[{\cal W}^{\alpha}\pi_{\alpha},\hspace{0.3em}\cdot \hspace{0.3em}]}&=& \int_0^{\infty} ds 
        e^{-s\left(p^2+m+[ {\cal W}^{\alpha}\pi_{\alpha},\hspace{0.3em}\cdot\hspace{0.3em} ]\right)} \nonumber \\
                &\hspace{-8em}=& \hspace{-5em}\int_0^{\infty}e^{-s\left(p^2+m\right)} 
          \left( 1-s[{\cal W}^{\alpha}\pi_{\alpha},\hspace{0.3em}\cdot\hspace{0.3em}]
                   +\frac{s^2}{2}\left( [{\cal W}^{\alpha}\pi_{\alpha},\hspace{0.3em}\cdot\hspace{0.3em} ] 
                          [{\cal W}^{\beta}\pi_{\beta},\hspace{0.3em} \cdot\hspace{0.3em}]\right) \right)\hspace{2em}.
 \label{pro}
\end{eqnarray} 
Since $\pi_{\alpha}$ is a Grassmann two-component spinor, the expansion stops at the second order. Since the fields are in the
 adjoint representation of $U(N)$, the action of ${\cal W}$ is through commutators. The propagators (\ref{pro}) can be 
represented as Fig.\ref{insertion_U_N}.
We can insert at most two $W^{\alpha}$ in each propagator. One important aspect of 
the propagator is the fact that the superfield ${\cal W}_{\alpha}$ is correlated with the fermionic momentum: ${\cal W}$ always 
appears with $\pi$. 

 Let $P$ be the number of propagators, $V$ be the number of vertices and $I$ be the number of index loops (for example,
$P=3$, $V=2$ and $I=3$ for the diagram $(a)$). Euler's 
theorem tells us that
\begin{equation}
   V-P+I=\chi,
   \label{Euler}
\end{equation}
where $\chi$ is the Euler number. The Feynman diagram also has some number $L$ of momentum loops. Using the relation
$L=P-V+1$, the above equation (\ref{Euler}) can be written as
\begin{equation}
   I-L=\chi -1.
 \label{Euler_2}
\end{equation}
In a diagram with $L$ loops we have to integrate over $2L$ Grassmannian momenta $\pi_{\alpha}$. Since these momenta necessarily
 appear in bilinears together with the background gauge field ${\cal W}^{\alpha}$, a diagram with $L$ loops will contribute 
precisely a factor of $({\cal W})^{2L}$ (with various possible gauge and spinor index contractions) to 
the effective superpotential. Since we are computing the superpotential for $S \sim tr(WW)$, $({\cal W})^{2L}$ must be 
arranged as follows
\begin{equation}
   tr(WW)tr(WW)\cdots tr(WW).
\end{equation}
This requires that the number of trace be greater than or equal to the momentum loop. Since the number of traces coincides with 
the number of index loops, the following relation must be satisfied
\begin{equation}
   I \geq L.
 \end{equation}
 Using the eq.(\ref{Euler_2}), we get the following constraint on the topology of graphs
 \begin{equation}
    \chi \geq 1.
 \end{equation}
 This means that in the case of the matter field in the adjoint of $U(N)$, we are concerned with $S^2$ graphs ($\chi =2$), that 
is planar graphs. Non-planar diagrams do not contribute to the glueball superpotential. It is worth noting that we do not have to
 take large $N$ limit to project out the planar diagrams. Planarity is an exact consequence of supersymmetry.
 
\begin{figure}[htbp]
   \begin{center}
   \includegraphics{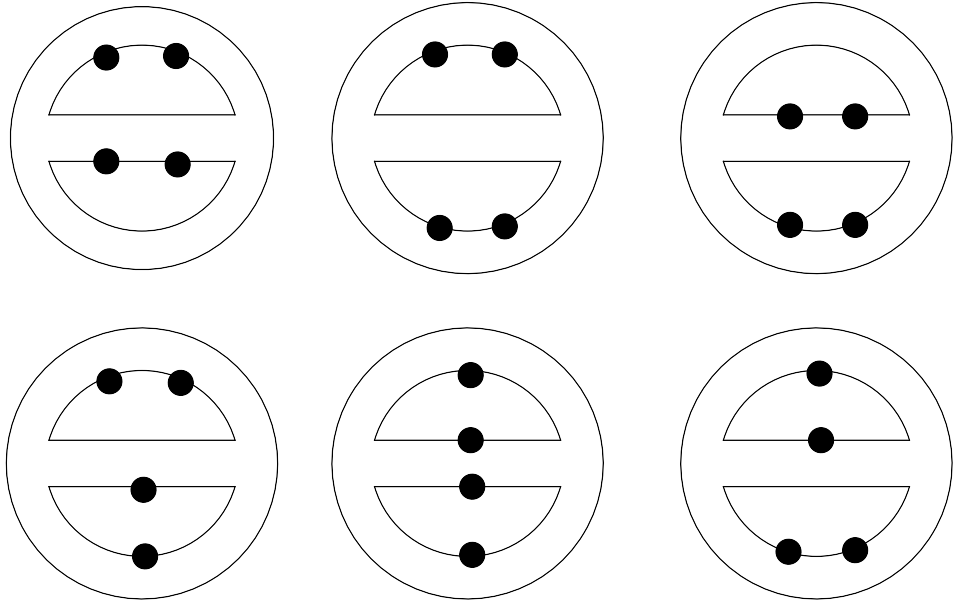}
   \caption{The "stop sign" diagrams with ${\cal W}$ insertions. We fixed the outer index loop to be free.}
   \label{insertion_U_N_vacuum}
   \end{center}
\end{figure}

 The planar diagrams at two-loop orders are diagram$(a)$ and $(b)$ depicted in Fig. \ref{vacuum_diagram_U_N}. Let us compute the 
"stop sign" diagram $(a)$. The amplitude is as follows
\begin{eqnarray}
   \frac{1}{2 !}\cdot 3 \cdot \left(\frac{g}{6}\right)^2 \int ds_1 ds_2 ds_3 
                \frac{d^4 p_1}{(2\pi )^4} \frac{d^4 p_2}{(2\pi )^4} d^2 \pi_1^2 d^2 \pi_2^2 
                 e^{-s_1(p_1^2+m+[{\cal W}^{\alpha},\hspace{0.3em} \cdot \hspace{0.3em}] \pi_{1\alpha})} \nonumber \\
               \times   e^{-s_2(p_2^2+m+[{\cal W}^{\alpha},\hspace{0.3em} \cdot \hspace{0.3em}] \pi_{2\alpha})}
                 e^{-s_3((-p_1-p_2)^2+m+[{\cal W}^{\alpha},\hspace{0.3em} \cdot \hspace{0.3em}] (-\pi_{1\alpha}-\pi_{2\alpha}))}.
\end{eqnarray}   
The bosonic integral is straightforward
\begin{eqnarray}
  Z_{boson} &=& \int \frac{d^4 p_1}{(2\pi )^4}\frac{d^4 p_2}{(2\pi )^4} e^{-s_1p_1^2-s_2p_2^2-s_3(p_1+p_2)^2}\nonumber \\
           &=& \frac{1}{(4\pi )^4}\frac{1}{(s_1s_2+s_2s_3+s_3s_1)^2}.
        \label{bosonic_integral}
\end{eqnarray}  
On the other hand, the computation of the fermionic integral:
\begin{equation}
  Z_{fermion}=\int d^2 \pi_1 d^2 \pi_2 e^{-s_1 [{\cal W}^{\alpha},\hspace{0.3em}\cdot \hspace{0.3em}]\pi_{1\alpha}}
                                       e^{-s_2 [{\cal W}^{\alpha},\hspace{0.3em}\cdot \hspace{0.3em}]\pi_{2\alpha}}
             e^{-s_3 [{\cal W}^{\alpha},\hspace{0.3em}\cdot \hspace{0.3em}](-\pi_{1\alpha}-\pi_{2\alpha})}
\end{equation}
 is more involved. In order to saturate the $\pi$ integral, four ${\cal W}$ have to be inserted at some point of the three
 index loops. We put two ${\cal W}$ on two index loops and leave the third loop without insertion(Fig.\ref{insertion_U_N_vacuum}) .
Summing up the all contributions we obtain
\begin{equation}
  Z_{fermion}=3NS^2(4\pi )^4(s_1 s_2 +s_2 s_3+s_3 s_1)^2,
\end{equation}
where the factor of $3$ counts the number of ways to pick one free (no ${\cal W}$ insertion) index loop out of 3 index loops.
$N$ comes from the trace over the free index loop: $tr(1)=N$. Comparing with the result of the bosonic integral
(\ref{bosonic_integral}), we observe the striking fact that the $s_i$ dependent part of the numerator and the denominator exactly
cancel:
\begin{equation}
   Z_{boson}Z_{fermion}=3 N S^2.
\end{equation}   
The result is independent of the Schwinger parameters $s_i$. Now the integral over $s_1, s_2$ and $s_3$ becomes trivial
\begin{equation}
  \int ds_1 ds_2 ds_3 e^{-m(s_1+s_2+s_3)}=\frac{1}{m^3}.
\end{equation}  
The final result for the stop sign diagram is thus
\begin{equation}
     \frac{1}{24}\frac{g^2}{m^3}3NS^2.
\end{equation}

 In general the amplitude corresponding to a planar diagram $\gamma$ can be written as
 \begin{equation}
    A^{\gamma}_{planar}=c_{\gamma}\int \prod_{i=1}^P ds_i e^{-s_i m}Z_{boson}Z_{fermion},
 \end{equation}   
 where $c_{\gamma}$ denotes the numerical factor. It is shown that the product of the bosonic momentum
 integral $Z_{boson}$ and the fermionic momentum integral $Z_{fermion}$ is independent of the Schwinger parameters $s_i$
 and given by
 \begin{equation}
    Z_{boson}Z_{fermion}=NIS^L. \hspace{2em}(I=L+1)
  \label{Z_U_N}
 \end{equation}   
 The integral over the Schwinger parameters are easily carried out
 \begin{equation}
    \int \prod_{i=1}^P d s_i e^{-s_i m}=\frac{1}{m^P}.
 \end{equation}   Thus, all that is left from the propagators is a contribution of $\frac{1}{m}$. This factor can be reproduced 
 by a mass term in the zero dimensional action $\frac{m}{2}tr(\Phi ^2)$. In this way the computation of glueball superpotential
 reduces to the evaluation of a zero dimensional field theory, i.e, matrix theory. 
 
   It was shown in \cite{Dijkgraaf:2002xd, Cachazo:2002ry} that the perturbative part of the effective superpotential $W_{eff}(S)$
of $U(N)$ gauge theory is related to the free energy of the matrix model whose potential equal to the tree level superpotential:
\begin{equation}
    W^{pert.}_{eff}(S)=N\frac{\partial {\cal F}_0(S)}{\partial S},
\end{equation} 
where ${\cal F}_0$ is the planar contribution to the free energy and the glueball-field $S$ is identified with the 't Hooft
coupling $gN$ of the matrix model. 

 We have investigated the perturbative part of the glueball superpotential so far. The full effective superpotential also includes the
Veneziano-Yankielowicz term:
\begin{equation}
   W_{eff}(S)=W_{VY}(S)+W^{pert.}_{eff}(S).
\end{equation} 
$W_{VY}$ is given by (\cite{Veneziano:1982ah})
\begin{equation}
    W_{VY} (S)=-h S\left(\log \left(\frac{S}{\Lambda}\right) -1 \right),
\end{equation}
where $h$ is the dual Coxeter number of the gauge group ($h =N$ for $U(N)$) and $\Lambda$ is the scale of the gauge theory.
$W_{VY}$ is the pure gauge part of the effective superpotential and is included "by hand" in our approach.

\subsection{$SO(N)$ gauge theory with an adjoint matter}

\begin{figure}[htbp]
 \begin{center}
   \includegraphics{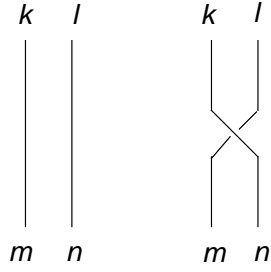}
   \caption{The free propagator of $\Phi$ in terms of 't Hooft double line notation.}
   \label{propagator_SU_N}
 \end{center}
\end{figure}

\begin{figure}[htbp]
  \begin{center}
     \includegraphics{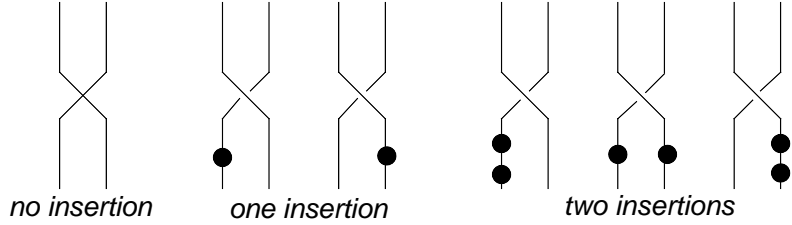}
     \caption{The twisted propagators with ${\cal W}$ insertions. The dots denote the ${\cal W}$ insertions.}
     \label{insertion_SO_N}
  \end{center}
\end{figure}

  The $SO(N)$ gauge theory with an adjoint matter was discussed in \cite{Ita:2002kx}. The adjoint fields of $SO(N)$ are
 antisymmetric $N \times N$ matrices:
\begin{equation}
      \Phi_{mn}=-\Phi_{nm}
\end{equation}
  and their free propagator in momentum space is proportional to the projector 
$P_{klmn}=\frac{1}{2}(\delta_{km}\delta_{ln}-\delta_{lm}\delta_{kn})$ \cite{Ita:2002kx, Kraus:2003jf}
\begin{equation}
  \langle \Phi_{kl}\Phi_{mn}\rangle  =\left( \frac{P}{p^2+m+[{\cal W}^{\alpha},\hspace{0.3 em}\cdot \hspace{0.3em}]
                        \pi_{\alpha}}\right)_{klmn}
                  = \left( \int_0^{\infty} ds e^{-s (p^2 +m+[{\cal W}^{\alpha},\hspace{0.3em}\cdot \hspace{0.3em}]\pi_{\alpha})}
                        \right)_{klmn}.
\end{equation}
For $SO(N)$ there are two-types of propagators: one is represented by parallel lines and the other is by crossed lines
(Fig.\ref{propagator_SU_N}).
The crossed lines with ${\cal W}$ insertions are represented in Fig.\ref{insertion_SO_N}.

As a consequence of the crossed lines, vacuum diagrams contain not only orientable Riemann surfaces but also non-orientable
surfaces. As discussed above, diagrams with $\chi \geq 1$ contribute to the glueball superpotential. Since $RP^2$ graphs meet the
requirement $\chi \geq 1$, they have non-vanishing contribution to the glueball superpotential. 

 It was shown in \cite{Ita:2002kx} that for the $RP_2$ graph the product of the bosonic integral and the fermionic integral 
 becomes
 \begin{equation}
  Z_{boson}Z_{fermion}=-4 \left( \frac{1}{2} \right)^P S^L.
  \label{Z_SO_N}
 \end{equation}
 As in the case of $S^2$ graphs, the Schwinger parameters cancel between bosons and fermions. This means that the computation
 of the glueball superpotential reduces to the evaluation of the zero-dimensional field theory.
 
  Using the above fact, Ita {\it et al} computed the glueball superpotential for the quartic interaction:
\begin{equation}
   W_{tree}(\Phi) = \frac{m}{2}tr(\Phi^2 ) +2g \ tr (\Phi^4)
\end{equation}
 and they obtained the following result
 \begin{equation}
  W^{pert.}_{eff}(S)= (N-2)[3S^2\left(\frac{g}{m^2}\right) +36 S^3 \left( \frac{g}{m^2}\right)^2+\cdots ].
 \end{equation}

\section{$G_2$ gauge theory with an adjoint matter}

  In this section we extend the analysis to $G_2$ gauge theory with an adjoint matter. $G_2$ differs from other Lie groups in that
  it has invariant three-rank tensor $f_{ijk}$, where $i,j$ and $k$ run from $1$ to $7$ \cite{Cvitanovic:1976am}. We 
  summarize the properties of $f_{ijk}$ in appendix.
  
\begin{figure}[htbp]
   \begin{center}
     \includegraphics{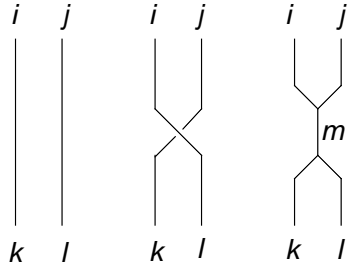}
     \caption{The three types of propagators.}
     \label{propagator_G_2}
   \end{center}
\end{figure}

\begin{figure}[htbp]
 \begin{center}
   \includegraphics{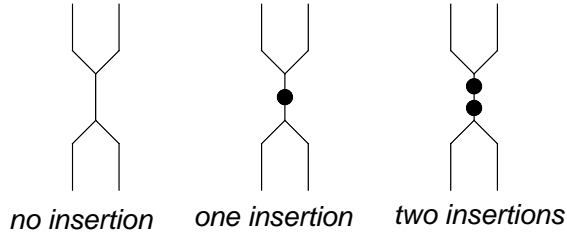}
   \caption{The propagators with ${\cal W}$ insertions}
   \label{insertion_G_2}
  \end{center}
\end{figure}

  The adjoint fields satisfy the following relations
\begin{equation}
   \Phi_{mn}=-\Phi_{nm}, \quad f_{ijk}\Phi_{jk}=0.
\end{equation}   
Their free propagator $\langle \Phi_{ij}\Phi_{kl}\rangle$ is proportional to the projector
\begin{equation}
  P_{ijkl}=\frac{1}{2}(\delta_{ik}\delta_{jl}-\delta_{il}\delta_{jk})-\frac{1}{6}f_{ijm}f_{mkl}.
  \label{projector}
\end{equation}
This projector is of course idempotent: $P=P^2$. The rule for multiplying four-index objects is
\begin{equation}
    (AB)_{ijkl}=\sum_{mn}A_{ijmn}B_{mnkl}.
\end{equation}
The presence of the three sorts of terms in the projector (\ref{projector}) means that in double line notation we have
three types of propagators, displayed in Fig.\ref{propagator_G_2}.
%When we take into account the background ${\cal W}$, 
The propagator in momentum space becomes
\begin{equation}
 \langle \Phi_{ij}\Phi_{kl}\rangle =\left(\frac{P}{p^2+m+[{\cal W}^{\alpha},\hspace{0.3em}\cdot \hspace{0.3em}]\pi_{\alpha}}
         \right)_{ijkl} = \left( \int_0^{\infty} ds e^{-s(p^2+m+[{\cal W}^{\alpha}, \hspace{0.3em}\cdot \hspace{0.3em}]\pi_{\alpha}
         )}P\right)_{ijkl}
\end{equation}
By expanding the fermionic part, we obtain
\begin{eqnarray}
  \left( e^{-s[ W^{\alpha}\pi_{\alpha},\hspace{0.3em}\cdot\hspace{0.3em} ]}P\right)_{ijkl}
          &=& P_{ijkl}-s\left([W^{\alpha}\pi_{\alpha},\hspace{0.3em}\cdot\hspace{0.3em} ]P\right)_{ijkl}
            +\frac{s^2}{2}\left( [W^{\alpha}\pi_{\alpha},\hspace{0.3em}\cdot\hspace{0.3em} ] 
                          [W^{\alpha}\pi_{\alpha},\hspace{0.3em} \cdot\hspace{0.3em}] P\right)_{ijkl}
       \nonumber  \end{eqnarray} \begin{eqnarray}
       &=& \frac{1}{2}\left( \delta_{ik}\delta_{jl}-\delta_{il}\delta_{jk}\right)-\frac{1}{6}f_{ijm}f_{mkl} \nonumber \\
       && +s \left( -\frac{1}{2}(W^{\alpha}\pi_{\alpha})_{ik}\delta_{lj}+\frac{1}{2}\delta_{ik}(W^{\alpha}\pi_{\alpha})_{lj}
          \right. \nonumber \\
       && \hspace{1.7em}+\frac{1}{2}(W^{\alpha}\pi_{\alpha})_{il}\delta_{kj}-\frac{1}{2}\delta_{il}(W^{\alpha}\pi_{\alpha})_{kj}\nonumber \\
       && \left. \hspace{1.7em}+\frac{1}{6}f_{ijka}(W^{\alpha}\pi_{\alpha})_{ab}f_{bkl}\right) \nonumber \\
       && +\frac{s^2}{2} \left( \frac{1}{2}(W^{\alpha}\pi_{\alpha}W^{\beta}\pi_{\beta})_{ik} \delta_{lj}
                                 -(W^{\alpha}\pi_{\alpha})_{ik}(W^{\beta}\pi_{\beta})_{lj}
                                 +\frac{1}{2}\delta_{ik}(W^{\alpha}\pi_{\alpha}W^{\beta}\pi_{\beta})_{lj} \right. \nonumber \\
                                &&\hspace{2.1em} -\frac{1}{2}(W^{\alpha}\pi_{\alpha}W^{\beta}\pi_{\beta})_{il}\delta_{kj}
                                 +(W^{\alpha}\pi_{\alpha})_{il}(W^{\beta}\pi_{\beta})_{kj}
                                 -\frac{1}{2}\delta_{il}(W^{\alpha}\pi_{\alpha}W^{\beta}\pi_{\beta})_{kj}\nonumber \\
                      &&\left.  \hspace{2.1em}-\frac{1}{6}f_{ija}(W^{\alpha}\pi_{\alpha}W^{\beta}\pi_{\beta})_{ab}f_{bkl}\right).
\end{eqnarray}
 We show the propagators including the  three-tensor $f$  with ${\cal W}$ insertions in Fig. \ref{insertion_G_2}.

\begin{figure}[htbp]
  \begin{center}
    \includegraphics{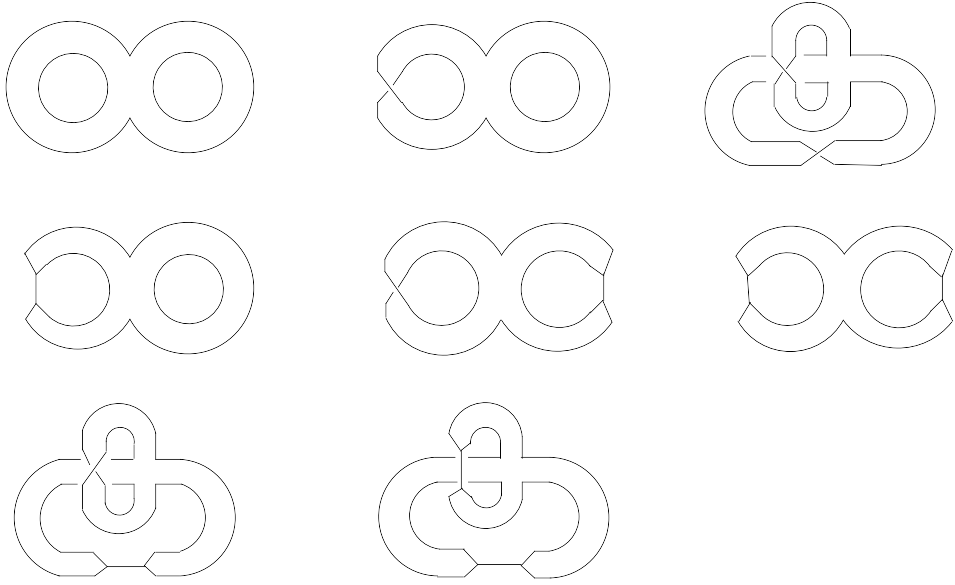}
    \caption{The ribbon Feynman diagrams contributing the glueball superpotential at two-loop level.}
    \label{two_loop_G_2}
  \end{center}
\end{figure}

Since the new type of propagator appears, the number of the diagrams contributing the glueball superpotential increases. 
For the quartic superpotential:
\begin{equation}
  W_{tree}(\Phi) = \frac{m}{2}tr (\Phi^2) +2g \ tr(\Phi^4),
\end{equation}  
we have to evaluate the diagrams sketched as in Fig.\ref{two_loop_G_2}   at two-loop order.
As in the case of $U(N)$ and $SO(N)$, we find that the product of the bosonic integral and the fermionic integral
is independent of the Schwinger parameters\footnote{At three-loop order, in addition to the glueball $S^3$, other gauge
 invariant quantities such as $ f_{ijk}f_{lmn}
{\cal W}^{\alpha}_{ia}({\cal W}_{\alpha})_{al}{\cal W}^{\beta}_{jb}({\cal W}_{\beta})_{bm}{\cal W}^{\gamma}_{kc}
({\cal W}_{\gamma})_{cn}$ appear. Since we are looking for the glueball superpotential, we neglect other gauge
 invariant quantities.}
\begin{equation}
    Z_{boson}Z_{fermion} =(const.)S^L.
    \label{Z_G_2}
\end{equation}
This means that the computation reduces to the evaluation of a zero-dimensional field theory. Although the cancellation
of the Schwinger parameters simplifies the calculation, the numerical factors of eq.(\ref{Z_G_2}) seem to obey no obvious rule
(see eq.(\ref{Z_U_N}) for $U(N)$ and eq.(\ref{Z_SO_N}) for $SO(N)$). The evaluation of the numerical factor must be done 
diagram by diagram.    

We obtain the following result: 
\begin{equation}
  W_{eff}(S)=4S\left( 1-\log\left( \frac{S}{\Lambda^3}\right) \right) +8\frac{g}{m^2}S^2+\frac{176}{3}\frac{g^2}{m^4}S^3+\cdots .
\label{final_result}
\end{equation}
The Veneziano-Yankielowicz term is included "by hand"($h=4$ for $G_2$). 

Finally we extremize $W(S)$ with respect to the $S$: $\frac{d W}{d S}=0$. The glueball $S$ acquire vacuum expectation 
values (glueball condensation):
\begin{equation}
   \langle S \rangle =\Lambda^3+4 \frac{g}{m^2}\Lambda^6 + 68\frac{g^2}{m^4}\Lambda^9 +\cdots,
   \label{gaugino_condensation}
\end{equation} 
spontaneously, breaking the chiral symmetries of the low energy effective gauge theory. Substituting 
eq.(\ref{gaugino_condensation}) into eq.(\ref{final_result}), in other words, integrating out $S$, we obtain
\begin{equation}
   W_{eff}=4\Lambda^3+8\frac{g}{m^2} \Lambda^6 +\frac{272}{3}\frac{g^2}{m^4}\Lambda^9 +\cdots.
   \label{eff}
\end{equation}   
The first terms of eq.(\ref{gaugino_condensation}) and eq.(\ref{eff}) are coming from pure gauge sector. On the other hand, 
the $\Lambda^6$ terms and $\Lambda^9$ terms
are coming from a matter field and depend on $W_{tree}(\Phi)$. Thus we obtain the non-perturbative correction to the gaugino
 condensation and effective Lagrangian. ($\Lambda^6$ terms and $\Lambda^9$ terms are the non-perturbative correction due to a 
fractional instanton of charge $\frac{1}{2}$ and $\frac{3}{4}$ respectively.) In this 
way, minimizing the effective superpotential turns a perturbative effect into the non-perturbative correction. It seems to be 
difficult or (nearly) impossible to obtain the eq.(\ref{gaugino_condensation}) and eq.(\ref{eff}) by traditional means.

\section{Conclusions and Discussion}

In this paper, following the diagrammatic approach of \cite{Dijkgraaf:2002xd}, we performed the perturbative computation of the
 glueball superpotential for $G_2$ gauge theory with an adjoint matter. The computation simplified because of the cancellation
between the bosonic momentum integral and the fermionic integral and reduced to the evaluation 
of zero-dimensional field theory. We evaluated $W(S)$ up to three-loop order. And then, by extremizing the glueball
 superpotential with respect to $S$, we gained the non-perturbative information about the gauge theory.

  Finally, we discuss an ambiguity at higher orders. In \cite{Kraus:2003jf}, Kraus and Shigemori calculated the glueball 
superpotential for $Sp(N)$ with an antisymmetric matter and found a discrepancy between the perturbative glueball superpotential
results and standard supersymmetric gauge theory results. The discrepancy showed up at order $h$, the dual Coxeter number. The 
reason is that $S^h$ is classically zero  and the perturbative computation beyond this order makes no sense.  
The classical relation $S^h =0$ was proven for $U(N)$ in \cite{Cachazo:2002ry}, for $SO(N)/Sp(N)$ in \cite{Witten:2003ye}, and
 for $G_2$ in \cite{Etingof:2003dd}.  
%The fact that classically $S^{h+n}=0$ for any non-negative $n$,
%implies that perturbatively it makes sense only to compute the glueball superpotentials for powers up to $h$. Beyond this, the 
%perturbative computation is ambiguous and UV dependent. The resolution of this ambiguity is discussed in \cite{Aganagic:2003xq}.
In the case of $G_2$ gauge theory, the perturbative computation is valid up to three loops. We evaluated the glueball 
superpotential up to this order. The resolution of the ambiguity at higher order is discussed in \cite{Aganagic:2003xq}.

\section*{Acknowledgments}
The author would like to thank Yoshihisa Kitazawa for useful discussions.

\setcounter{section}{0}
\renewcommand{\thesection}{A}
\section*{Appendix  \hspace{0.3em} $G_2$ invariant three-tensor $f_{ijk}$}

$G_2$ has invariant three-tensor $f_{ijk}$ ($i,j,k$ run from 1 to 7)\cite{Cvitanovic:1976am,Pesando:1995bq,Giddings:1995ns}. 
The following identities are satisfied.
\begin{equation}
  f_{iln}f_{jkn}+f_{jln}f_{ikn}=2\delta_{ij}\delta_{lk}-\delta_{il}\delta_{jk}-\delta_{ik}\delta_{jl}
\end{equation}
\begin{equation}
  f_{iln}f_{jln}=6\delta_{ij}
\end{equation}
\begin{eqnarray}
   f_{ijm}f_{mkn}f_{nlp}&=& + \delta_{ik}f_{jlp}+\delta_{pk}f_{ijl}+\delta_{ip}f_{jkl}+\delta_{jl}f_{ikp}\nonumber \\
           &&-\delta_{jk}f_{ilp}-\delta_{lk}f_{ijp}-\delta_{il}f_{jkp}-\delta_{jp}f_{ikl}
\end{eqnarray}  
\begin{equation}
   f_{ijm}f_{mkn}f_{nli}=3f_{jkl}
\end{equation}
\begin{eqnarray}
f_{aib}f_{bjc}f_{ckd}f_{dla}=5\delta_{ij}\delta_{kl}-4\delta_{ik}\delta_{jk}+5\delta_{il}\delta_{jk}
\end{eqnarray}
Since $f_{ijk}$ is an invariant tensor, the following identity is satisfied\cite{Cvitanovic:1976am}
\begin{equation}
  (T^a)_{im}f_{mjk}+(T^a)_{jm} f_{imk}+(T^a)_{km}f_{ijm}=0,
\end{equation}
where $T^a (a=1,\cdots , 14)$ is the generator of $G_2$.
By multiplying $W^{a\alpha}$, we obtain
\begin{equation}
 (W^{\alpha})_{im}f_{mjk}+(W^{\alpha})_{jm}f_{imk}+(W^{\alpha})_{km}f_{ijm}=0.
\end{equation}

\end{document}